\documentclass[10pt,twocolumn]{article} 
\usepackage{arxivStyle}
\usepackage{times}
\usepackage{graphicx}
\usepackage{amssymb}
\usepackage{url,hyperref}
\usepackage{float}
\usepackage{array}
\usepackage{tabularx}
\usepackage{booktabs}
\usepackage{amsmath}
\usepackage{placeins}

\begin{document}

\title{Deep Learning-Based Segmentation of Tumors in PET/CT Volumes: Benchmark of Different Architectures and Training Strategies}

\author{Monika Górka $^{1,\dagger}$, Daniel Jaworek $^{1,\dagger}$, and Marek Wodzinski $^{1,2}$ \\
\\
$^{1}$ AGH University of Krakow, Department of Measurement and Electronics \\ Krakow, Poland wodzinski@agh.edu.pl \\
$^{2}$ University of Applied Sciences Western Switzerland (HES-SO Valais) \\  Information Systems Institute, Sierre, Switzerland \\
$^{\dagger}$ These authors contributed equally to this work.
}

\maketitle
\thispagestyle{empty}

\begin{abstract}
Cancer is one of the leading causes of death globally, and early diagnosis is crucial for patient survival. Deep learning algorithms have great potential for automatic cancer analysis. Artificial intelligence has achieved high performance in recognizing and segmenting single lesions. However, diagnosing multiple lesions remains a challenge. This study examines and compares various neural network architectures and training strategies for automatically segmentation of cancer lesions using PET/CT images from the head, neck, and whole body. The authors analyzed datasets from the AutoPET and HECKTOR challenges, exploring popular single-step segmentation architectures and presenting a two-step approach. The results indicate that the V-Net and nnU-Net models were the most effective for their respective datasets. The results for the HECKTOR dataset ranged from 0.75 to 0.76 for the aggregated Dice coefficient.  Eliminating cancer-free cases from the AutoPET dataset was found to improve the performance of most models. In the case of AutoPET data, the average segmentation efficiency after training only on images containing cancer lesions increased from 0.55 to 0.66 for the classic Dice coefficient and from 0.65 to 0.73 for the aggregated Dice coefficient. The research demonstrates the potential of artificial intelligence in precise oncological diagnostics and may contribute to the development of more targeted and effective cancer assessment techniques

\end{abstract}

\section{Introduction}
\subsection{Overview}

Globally, cancer is acknowledged as one of the primary sources of premature mortality, alongside cardiovascular diseases, and presents a formidable global health challenge. It is predicted that by the end of the century, cancer will overtake cardiovascular diseases, becoming the primary cause of premature mortality in most nations across the world \cite{bray2021ever}. Early detection of cancer lesions in patients is crucial for improving survival rates.  The prognosis and treatment options depend on the location and stage of the lesions \cite{munir2019cancer, chow2020head}. Computed tomography (CT) and positron emission tomography (PET) are playing a key role in tumor analysis, providing valuable information about the location, anatomy and stage of the tumor. Manual analysis and interpretation of medical images is a time-consuming and error-prone process that may lead to a lack of consistency in conclusions between different specialists \cite{oreiller2022head, rebaud2022simplicity}. Therefore, it is extremely important to continuously develop deep learning methods to incorporate them into healthcare to accelerate and improve diagnostic processes and analysis of biological data, which will increase the effectiveness of detecting and understanding cancer mechanisms \cite{tran2021deep}. Currently, despite the high effectiveness of deep learning methods in detecting and segmenting single lesions, the task of diagnosing multiple lesions is still much more difficult due to low variability between lesions or too large a number of lesions \cite{jiang2023review}. Another significant challenge is the detection of small-volume tumors at an early stage of their development. In this situation, the PET/CT imaging technique plays an invaluable role, enabling the localization of even minimal cancer lesions. Moreover, the integration of methods based on deep learning in the analysis of data obtained from PET/CT imaging can significantly increase the efficiency of detection of early-stage small-volume tumors. This, in turn, has a direct impact on increasing the therapeutic potential and improving the life prognosis of cancer patients.

\subsection{Related Work}

Many publications have explored the application of deep learning techniques in automated cancer analysis. Our study focused on two issues: automatic analysis of head and neck cancers and automatic analysis of whole-body cancers. Authors of \cite{wang2021deep} reviewed the literature on deep learning techniques in multiomics analysis and diagnosis of head and neck cancers, with a focus on convolutional neural networks (CNNs). In \cite{illimoottil2023recent} advanced methods of deep learning were introduced by the authors, such as convolutional autoencoders, generative adversarial networks (GANs), and transformers, to perform automatic analysis of head and neck cancers for multiple tasks. In \cite{li2022application} the applications of Deep Learning Radiomics (DLR) technology in predicting head and neck cancer were described, and it was concluded that DLR has significant potential for segmenting regions of interest (ROI), extracting and fusing features, and constructing models using PET/CT volumes. Article \cite{franzese2023enhancing} reviews the use of artificial intelligence to enhance the radiotherapy workflow for head and neck cancer, focusing on segmentation, planning, and delivery. Study \cite{wang2024comparison} focuses on the comparison of deep neural networks for automatic localization and segmentation of head and neck cancers based on PET/CT images. A study \cite{shiri2024information} aimed to assess the performance of PET and CT image fusion for gross tumor volume segmentations of head and neck cancers utilizing conventional, deep learning, and output-level voting-based fusions. In \cite{zhou2017deep} an approach for automatically segmenting multiple organs in three-dimensional computed tomography (CT) images using deep learning was proposed. In study \cite{bauer2012automated}, researchers examined the use of a cascade convolutional neural network architecture to automatically identify and segment tumors in whole-body FDG-PET images. Study \cite{jiang2023review} provided a review of deep learning algorithms that can detect multifocal neoplastic cancer lesions in the human body, including scenarios that involve whole-body imaging. The study \cite{shi2022deep} presents an unconventional approach to precise initialization of organs-at-risk and tumors. The authors proposed a framework implementing a cascade coarse-to-fine segmentation with adaptive module for both small and large organs. The paper \cite{xue2024multi} presents an overview of the latest deep learning-based multimodal tumor segmentation methods. Public datasets, evaluation methods and multimodal data processing are presented, as well as typical deep learning network structures, techniques and methods of multimodal image fusion used in various tumor segmentation tasks. Publication \cite{gatidis2023autopet} summarizes the AutoPET Grand Challenge 2022 and describes the whole-body tumor segmentation methods proposed by the participants. Publications \cite{andrearczyk2020overview,andrearczyk2021overview,andrearczyk2022hecktor} provide summaries of the HECKTOR challenges from 2020, 2021, and 2020, respectively, along with a description of the head and neck tumor segmentation methods employed by the participants.

\subsection{Contribution}

This study contributes to the field of diagnostic imaging in oncology by analyzing, modifying, and comparing advanced deep learning algorithms for cancer lesion segmentation from PET/CT images. The study evaluates the performance of various neural network architectures commonly used in medical image analysis. The careful comparison of these models enabled the identification of the most effective architectures for specific diagnostic applications, both for whole-body images and specific head and neck regions. Furthermore, emphasis was placed on the two-stage segmentation methodology, which, although less common in the literature, shows promising results compared to traditional one-stage segmentation. Additionally, the study examines the effects of excluding non-cancer cases from the training dataset, which is a novel and counter-intuitive method for optimizing deep learning models. Consequently, this research offers valuable insights into selecting and configuring neural network models to enhance their diagnostic imaging capabilities. These findings have significant implications for the development of more advanced and accurate diagnostic tools in oncology. This could lead to better clinical outcomes and increased survival rates for cancer patients.

\section{Materials and Methods}

\subsection{Overview}

To achieve satisfactory results in segmentation tasks, we focused on tools for effectively extracting features from medical images, despite the limited available data. The common small dataset problem arises from the challenge of obtaining large, well-labeled, and high-quality datasets. The efficacy of deep learning approaches is intricately connected to the size and quality of the available dataset \cite{wang2021deep, illimoottil2023recent}. Our research required tools capable of effectively segmenting a variety of tumor types with different anatomical structures located in different parts of the patient's body. In our study, we used two different datasets. The first one came from the HECKTOR (Head and Neck Tumor) Grand Challenge, and the second one came from the AutoPET Grand Challenge held at MICCAI (Medical Image Computing and Computer Assisted Intervention 2022). We used the UNet, VNet, and UNETR models to perform segmentation on the HECKTOR dataset, and the UNet, UNETR, and nnUnet models to perform segmentation on the AutoPET dataset.

\subsection{U-Net}

U-Net's fully connected layered encoder-decoder architecture has revolutionized semantic segmentation. The skip connections between the encoder and decoder are the essential component that enables the efficient concatenation of feature maps. Symmetric upsampling and downsampling processes significantly improve localization and segmentation accuracy. 
U-Net's flexibility, adaptability, and impressive performance for a variety of medical imaging modalities with minimal training data have led to its popularity. Consequently, this architecture has been modified for use in various challenges. For this reason, we decided to use the U-Net architecture from the MONAI library, which is a set of open-source, freely available collaborative frameworks built for accelerating research and clinical collaboration in medical imaging \cite{jiang2023review, azad2022medical, mo2022review, malhotra2022deep, conze2023current, ronneberger2015u}. To adapt the architecture to our datasets we used the standard version of the U-Net and modified its parameters without affecting the implementation of the model. Our modification included two input channels corresponding to the two modalities used: CT and PET. The network contained five layers with two convolutional units, with a kernel size of 3x3x3, where each coding layer contained residual units with two convolutional blocks and residual connections. The network was implemented with a PReLU (Parametric Rectified Linear Unit) activation function, an extension of ReLU, and a transposed convolution operation in the decoder, enabling a gradual reduction in the number of channels and an increase in spatial dimensions. For the AutoPET dataset, the network returns a three-dimensional binary mask indicating the detected tumor areas, while for the HECKTOR dataset, the network returns two binary masks - each mask corresponding to one of the two types of tumors present in the images.

\subsection{UNETR}

Transformers, first introduced in 2017 for natural language processing, have become increasingly important in medical image segmentation. Vision Transformers (ViT) offer a critical method for dividing images into fragments, which are processed by a transformer with a central multi-head auto-attention (MSA) layer. Models like TransU-Net and UNETR, which use transformers, enable more accurate image analysis by capturing long-range relationships and complex contexts, which is important for precise segmentation of medical images. Our study aimed to evaluate transformer effectiveness utilizing the UNETR model, which combines the transformer architecture with U-Net elements. This model divides the image into equal, non-overlapping fragments and projects them into the embedding space using a linear layer. Next, this sequence is augmented with positional embeddings and serves as input to the transformer model. Tensors are obtained at each resolution level of this model and then connected in a manner analogous to the standard U-net architecture. At the last stage, a convolutional layer, alongside a softmax activation function, is used \cite{conze2023current, xiao2023transformers, thisanke2023semantic, monai_unetr}. In our study, we used the UNETR model from the MONAI library. Like the UNet model, the network takes a two-channel input and returns one binary mask in the case of the AutoPET set and two binary masks in the case of the HECKTOR set.

\subsection{V-Net}
V-Net is a convolutional neural network specifically designed for processing volumetric medical data. The network architecture was inspired by the U-Net architecture and also consists of encoder and decoder parts that effectively extract image features and details, enabling precise segmentation \cite{milletari2016v}. The implementation we used from the MONAI library consists of 5 layers with volumetric convolution, ELU (Exponential Linear Unit) activation function and dropout in the last two layers. We applied the V-Net only to the HECKTOR dataset. The network takes a two-channel image as input and returns two binary masks.

\subsection{nnU-Net}

nnU-Net is currently one of the most prevalent tools for semantic segmentation in various scientific fields. It has achieved top results in several open challenges in medical imaging. nnU-net is based not only on the proper preparation of the neural network, but also on the automatic configuration of pre- and post-processing. Its decision-making process is based on a set of parameters, rules and empirical decisions. The framework generates a "dataset fingerprint" by examining the provided training examples, which is then used to develop appropriate configurations for datasets: 2d,  3d\_fullres, 3d\_lowres and 3d\_cascade\_fullres. Due to the automatic parameter adjustment and preprocessing, the aforementioned data preprocessing activities were not required. During training of the nnU-Net datasets, the 5-fold cross-validation method was employed. This method facilitated the evaluation of different configurations and resulted in the selection of the most optimal segmentation for a given problem \cite{isensee2021nnu, nnUNetUNETR2023}.

\subsection{Datasets}

We used two datasets in our study. The first one came from the HECKTOR (Head and Neck Tumor) Grand Challenge, and the second one came from the AutoPET Grand Challenge held at MICCAI (Medical Image Computing and Computer Assisted Intervention 2022) \cite{gatidis2023autopet, andrearczyk2022hecktor}. 

HECKTOR dataset includes 3D scans of the head and neck area from PET and CT.  For training cases, segmentation masks with marked tumor areas are additionally included. The masks present two structures: primary Gross Tumor Volumes (GTVp) and nodal Gross Tumor Volumes (GTVn). Data come from 9 centers situated in Canada, Switzerland, France, and the United States. The entire dataset contains 845 cases, divided into 524 training cases and 356 test cases \cite{andrearczyk2022hecktor, hecktorwebpage}. Only data from the training set was used to conduct the experiments due to the lack of access to segmentation masks for the test set.

The AutoPET dataset contains FDG-PET/CT images that cover the whole body from the base of the skull to the mid-thigh. The dataset include 1014 images obtained from 900 patients who took part in a prospective clinical trial conducted between 2014 and 2018 at the University Hospital of Tübingen. Among all the images, 501 were positive cases, indicating cancerous lesions that displayed enhanced metabolic activity of the FDG factor. This group comprises patients with non-small cell lung cancer (NSCLC), malignant lymphoma, or melanoma. The remaining 513 images exhibit negative cases in which no metabolically active cancer has been detected \cite{gatidis2022whole}.

\subsection{Experimental Setup}

The first step was to prepare the data. Preprocessing of the HECKTOR dataset included resampling to a common resolution and then cropping to the head and neck area to the size of 192x192x192. Images were cropped in the sagittal and coronal planes from the center of the image and in the transverse plane from the top of the image. The AutoPET data has already been resampled to a common resolution by the challenge organizers. The following preprocessing steps were the same for both data sets. The CT image intensity was cut to the range [-1024, 1024] Hounsfield Units, since this range covers most anatomical structures of the human body, and then rescaled to [-1, 1]. The PET image intensity was also rescaled to the range of [-1, 1]. The PET and CT images were then superimposed to result in a single two-channel image. For the HECKTOR dataset, each segmentation mask was divided into two binary masks. One mask corresponded to the GTVp areas, and the other mask corresponded to the GTVn areas. Cases with only one registered structure had one blank mask. In the case of AutoPET data, the original masks were already binary. There was no need to split them.

Data augmentations were utilized in all experiments to minimize overfitting. Augmentation methods included random flipping, rotation, scaling, translation, gaussian noise, and random gamma.

In the case of AutoPET data, tumors were located in different areas of the images, and there were discrepancies in size across images. To ensure both full resolution and high data quality while maintaining computational optimality, we implemented a patch-based approach for the U-Net model. This involved training the model on subsets of the full image, thus enabling resolution to be maintained. During the experiments, we utilized a modified Label Sampler that employed dynamic adjustment of a parameter responsible for controlling class sampling probability in  images \cite{perez2021torchio}. Patch size was standardized to 96x96x96, and 80 samples per image volume were taken. For the UNETR architecture, we decided to adopt a standard training method for all images due to hardware restrictions and the intricacy of the transformer structure. As a result, all images were cropped to 192x192x224.

Another challenge with AutoPET data, as mentioned in Section 2.6, is that approximately half of the images represent negative cases without cancerous lesions. Consequently, the impact of these cases on the quality of segmentation was evaluated through an experiment solely utilizing data featuring positive controls. Non-cancerous images were removed from the dataset. Comparative segmentation was then performed in the corresponding configurations for U-Net and UNETR.

\begin{figure}[!h]
    \centering
    \includegraphics[width=1\linewidth]{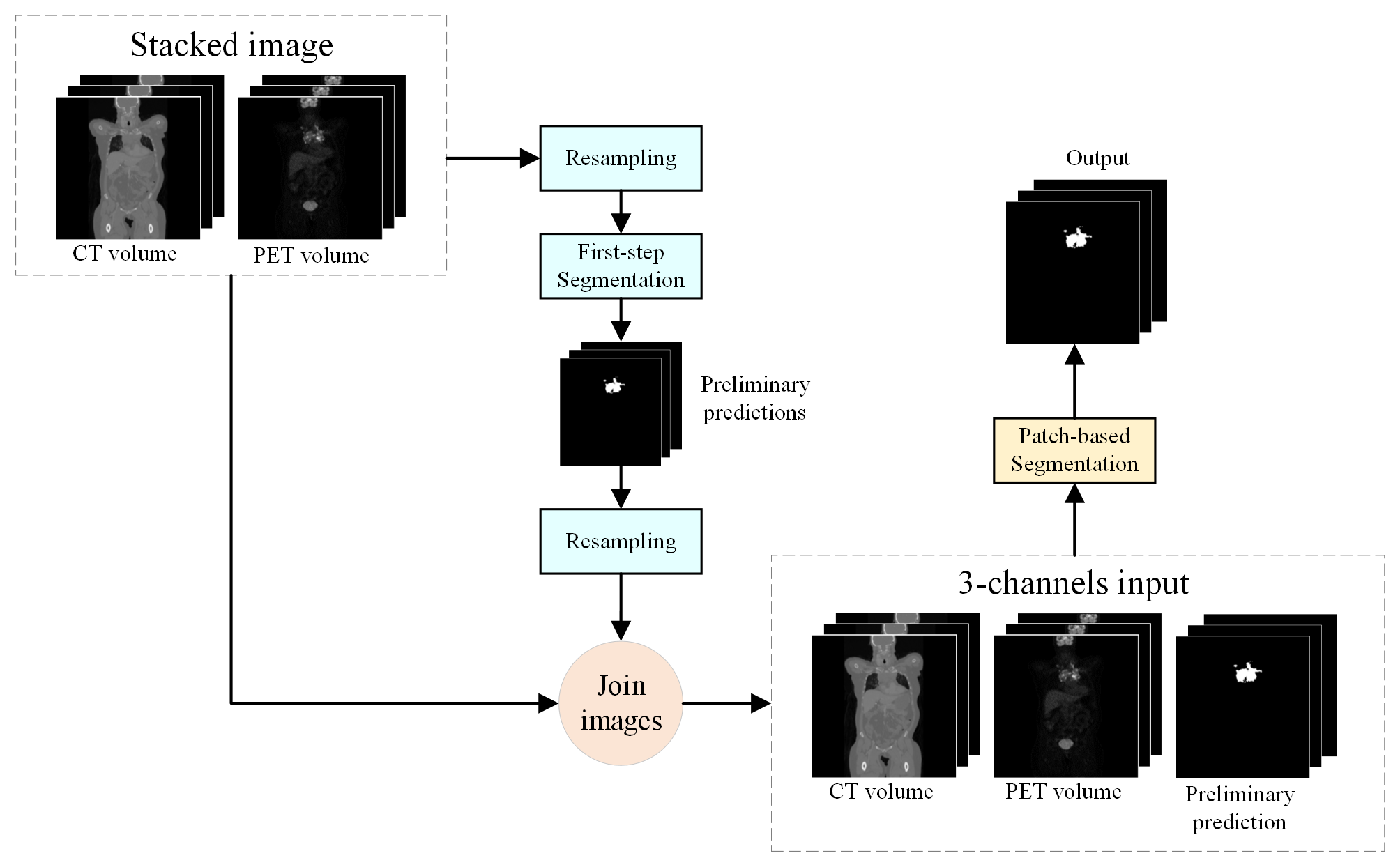}
    \caption{Diagram illustrating a implementation of two-step segmentation
process with three-channel aproach.}
    \label{fig:3channels}
\end{figure}

In order to improve the results of the U-Net model, an additional segmentation step was performed, which included first identifying potential tumor areas. Initially, a strategy was implemented to train the model on reduced dimension and resolution data. The data underwent processing via the applied dynamic resampling strategy, with input images being scaled to 192x192x192. As the initial model, the U-Net architecture with the original parameter configuration was applied. Following the model's initial training, segmentation mask predictions were generated across all data sets. The masks were reinserted into the spatial context of their corresponding source images and added as a third input channel to the U-Net model using the described configuration.

All models and configurations described above were trained using the Adam optimizer and Cosine Annealing with Warm Restarts learning rate scheduler, which cyclically lowers the value of the learning rate and then resets it to the initial value to prevent the model from getting stuck in the local minimum gradient during training. The loss function used was DiceFocalLoss. This function returns the weighted sum of Dice Loss and Focal Loss.

\section{Results}

\newcolumntype{C}{>{\centering\arraybackslash}X}
The effectiveness of the models was assessed using commonly used metrics for image segmentation tasks. The results were evaluated using the dice similarity coefficient (DSC), which is defined by the following formula:
\begin{equation}
    DSC\left(A,B\right) = \frac{2\left|{A \cap B}\right|}{\left|{A}\right| + \left|{B}\right|},
\end{equation}

where $A$ and $B$ are two sets whose similarity should be assessed. It was decided to use the aggregated Dice coefficient as the second metric. In the context of image segmentation, there may be two binary images representing regions of interest. The aggregated Dice coefficient provides information on the overall segmentation quality for all regions of interest, which allows for a comprehensive assessment of the algorithm's performance.

\begin{equation}
    DSC_{agg} = \frac{2\sum_{i}\left|{A_{i} \cap B_{i}}\right|}{\sum_{i}\left|{A_{i}}\right| + \left|{B_{i}}\right|},
    \label{eq:dcagg}
\end{equation}

where $A_i$ and $B_i$ are the ground truth and prediction for the image i in the test set.

\subsection{AutoPET}

As part of our study, we conducted a comparative analysis of the effectiveness of three different one-step segmentation methods. The experiment utilized three models: U-Net architecture, UNETR, and nnU-Net, a dedicated framework for image segmentation. The models were trained on two datasets. The first dataset contained all images of the initial dataset, while the second dataset was reduced by negative control cases in which there was no case of cancer. The second part of the study was to perform two-step segmentation on the U-Net model. The models' effectiveness was evaluated using two indicators: the Dice similarity coefficient and the aggregated Dice similarity coefficient. The obtained results are presented in Table~\ref{tab:autopet_tabela}. The evaluation was performed on a separate test set from the original dataset included images with cancerous lesions. Exemplary visualizations of the obtained results are shown in Figure~\ref{fig:3channels}.

\begin{table}[!h]
\centering
\caption{Presentation of the results of various segmentation methods.}
\begin{tabularx}{\linewidth}{C C C C C C}
\toprule
&& \multicolumn{2}{c}{\textbf{Full dataset}} & \multicolumn{2}{c}{\textbf{Tumor-only dataset}} \\ 
\midrule
& & \textbf{DSC ($\uparrow$)} & \textbf{DSC\textsubscript{agg} ($\uparrow$)} & \textbf{DSC ($\uparrow$)} & \textbf{DSC\textsubscript{agg} ($\uparrow$)} \\
\midrule
& U-Net & $\mathrm{0.58}$ & $\mathrm{0.64}$ & $\mathrm{\textbf{0.74}}$ & $\mathrm{0.80}$ \\
\textbf{One-step}& nnU-Net  & $\mathrm{\textbf{0.69}}$ & $\mathrm{\textbf{0.80}}$ & $\mathrm{\textbf{0.74}}$ & $\mathrm{\textbf{0.84}}$ \\ 
& UNETR & $\mathrm{0.39}$ & $\mathrm{0.51}$ & $\mathrm{0.51}$ & $\mathrm{0.54}$ \\ 
\midrule
\textbf{Two-steps} & U-Net  & $\mathrm{0.60}$ & $\mathrm{0.73}$ & - & - \\ 
\bottomrule
\end{tabularx}
\label{tab:autopet_tabela}
\end{table}

During the statistical analysis, we focused on analyzing the significance of differences between a specific model trained on the full dataset and the set containing only cancer data. To determine the statistical significance of differences in results for models trained on different datasets, a p-value was calculated. The Wilcoxon test \cite{mccrum2008correct} was used to calculate the p-value with the null hypothesis (H0) that the differences in results between the analyzed models are not statistically significant at the 0.05 significance level. 

The obtained p-values are presented in Table~\ref{tab:autopet_tabela_p}. P values below 0.00001 were approximated to 0.

\begin{table}[!h]
\centering
\caption{Presentation of the p-values for pairs of all models.}
\label{tab:autopet_tabela_p}
\begin{tabularx}{\linewidth}{C C C}
\toprule
\textbf{Full dataset} & \textbf{Tumor-only dataset} & \textbf{p-value} \\
\midrule
U-Net & U-Net  & \textasciitilde{$0$} \\
nnU-Net & nnU-Net  & $\mathrm{0.0031}$ \\
UNETR  & UNETR  & $\mathrm{0.0017}$ \\
\bottomrule
\end{tabularx}
\end{table}

\begin{figure}[H]
    \centering
    \includegraphics[width=0.9\linewidth]{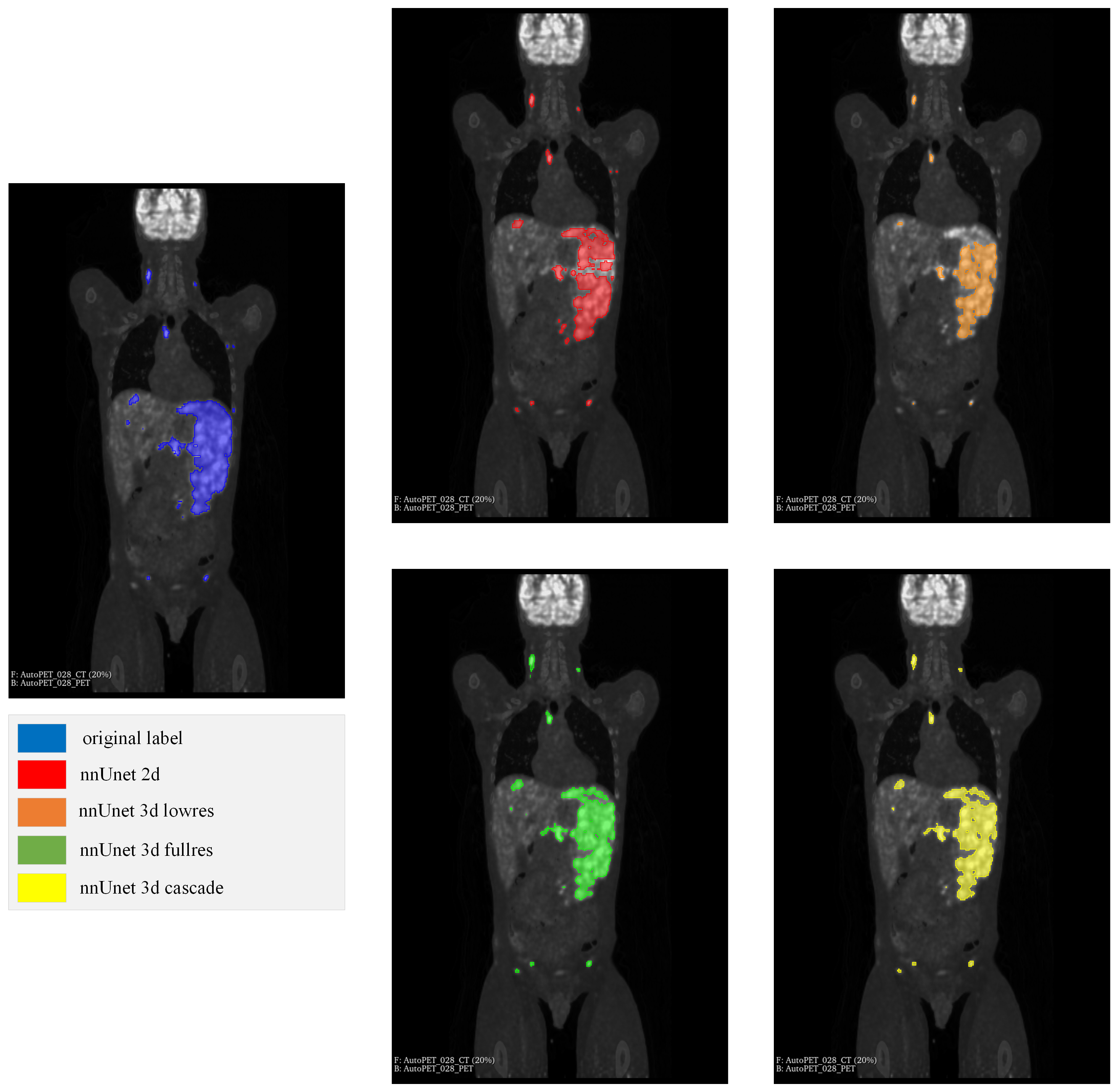}
    \caption{Visualization of the most accurate segmentation (DSC: 0.94) for
various nnU-Net configurations trained on the tumor-only dataset. The ground
truth tumor is marked in blue, while the model predictions are marked in red,
orange, green, and yellow.}
    \label{fig:3channels}
\end{figure}

\subsection{HECKTOR}

Three deep neural network architectures were compared on the HECKTOR dataset: U-Net, UNETR, and V-Net. The performance of the models was evaluated on a test set that was previously separated from the training set. Before the evaluation, the generated predictions were put back into the full-dimensional spatial context of the original corresponding CT image. Model performance was assessed using the aggregated Dice similarity coefficient. The script provided by the organizers of the HECKTOR challenge was used for evaluation \cite{evaluation}. The results are shown in Table~\ref{tab:hecktor_results}.

\begin{table}[!h]
\centering
\caption{ Segmentation results on the HECKTOR test set for models trained on cropped
images.}
\begin{tabularx}{\linewidth}{C C C C}
\toprule
& \textbf{DSC\textsubscript{agg}GTVp ($\uparrow$)} & \textbf{DSC\textsubscript{agg}GTVn ($\uparrow$)} & \textbf{mean DSC\textsubscript{agg}($\uparrow$)} \\
\midrule
U-Net & $\mathrm{0.79}$ & $\mathrm{0.72}$ & $\mathrm{0.76}$ \\
V-Net & $\mathrm{0.80}$ & $\mathrm{0.72}$ & $\mathrm{0.76}$ \\
UNETR & $\mathrm{0.79}$  & $\mathrm{0.69}$ & $\mathrm{0.74}$ \\
\bottomrule
\end{tabularx}
\label{tab:hecktor_results}
\end{table}

To determine the statistical significance of the disparities in the test set outcomes among the models, the p-value was calculated. The metric used to evaluate the models (DSC\textsubscript{agg}) is a global metric, calculated for the entire test set instead of for individual samples. Conventional statistical tests require multiple observations, and global metrics cannot directly satisfy this requirement. Therefore, bootstrap resampling \cite{efron1994introduction, noguchi2011bootstrap} was used. The test set was repeatedly sampled with replacement to generate new test sets. DSC\textsubscript{agg} was calculated for each drawn sample, allowing for the estimation of metric distributions across models.

As with the AutoPET data, the p-value was calculated using the Wilcoxon test with a null hypothesis that assumes no statistically significant differences in results between the analyzed models at a significance level of 0.05. The resultant p-values have been presented in Table 4.5. P-values below 0.00001 have been approximated to 0.

{
\newcolumntype{M}[1]{>{\centering\arraybackslash}m{#1}}

\renewcommand{\arraystretch}{2}

\begin{table}[!h]
\centering
\caption{Matrix displaying the p-values for pairs of segmentation models.}
\label{matrix:p_value_seg}
\[
\begin{array}{M{1.3cm}|M{1.3cm}|M{1.3cm}|M{1.3cm}}
 & \rotatebox[origin=c]{90}{U-Net} & \rotatebox[origin=c]{90}{V-Net} & \rotatebox[origin=c]{90}{UNETR}\\ \hline
U-Net & \text{$-$} & \textasciitilde{$0$} & \textasciitilde{$0$}\\ \hline
V-Net  & \textasciitilde{$0$} & \text{$-$} & \textasciitilde{$0$}\\ \hline
UNETR  & \textasciitilde{$0$} & \textasciitilde{$0$} & \text{$-$}
\end{array}
\]
\end{table}
}

\begin{figure}[!h]
  \centering
    \includegraphics[width=1.0\linewidth]{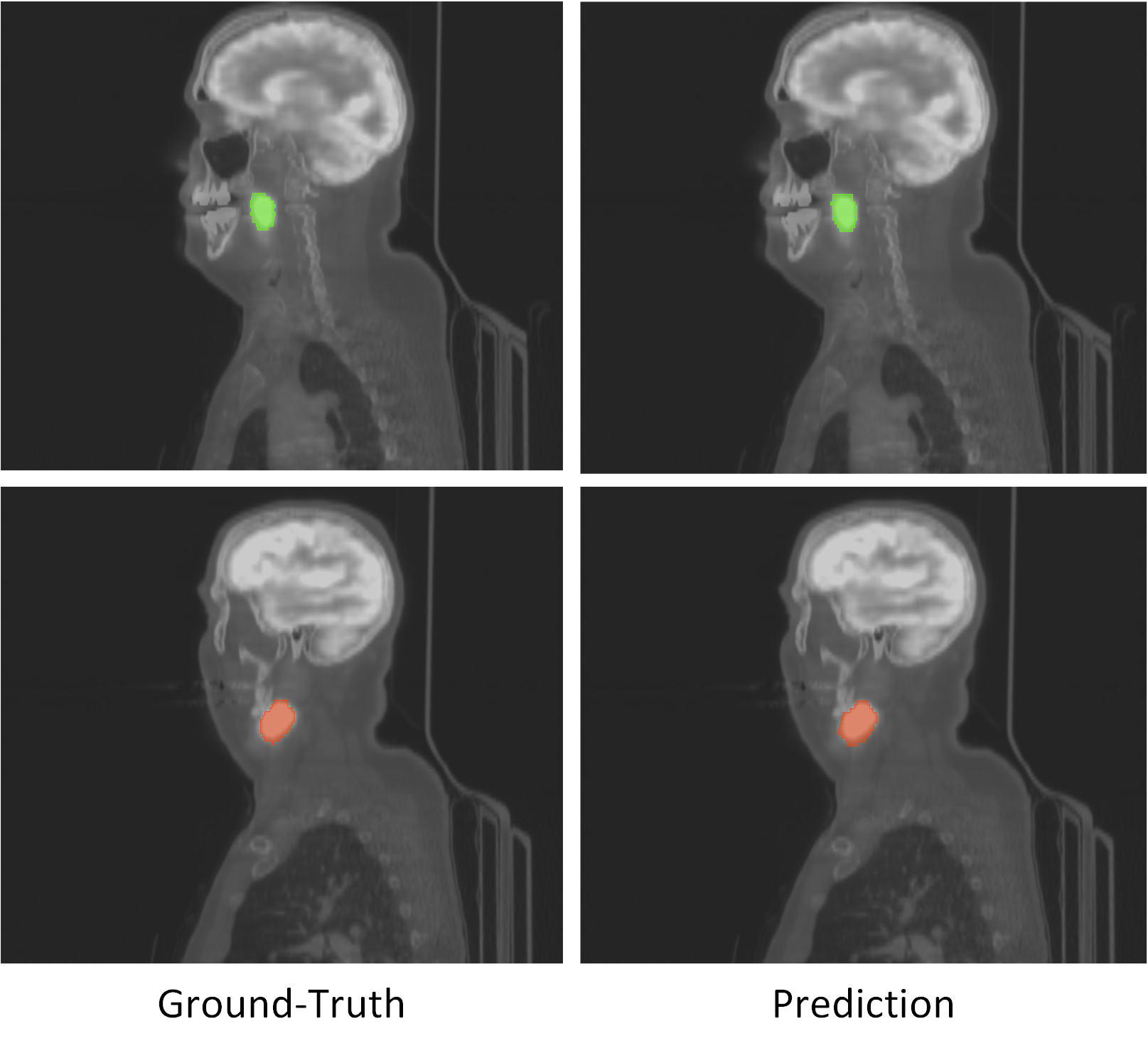}
  \caption{Exemplary visualizations of the obtained results from the HECKTOR dataset.}
  \label{fig:hecktor_viz}
\end{figure}

The models were evaluated using a test set extracted from the training set. This is distinct from the data on which the results of the HECKTOR challenge were evaluated. The results were submitted to the challenge to obtain the results for the HECKTOR test set, in order to be able to compare the obtained results with those of the challenge participants. Due to the limited number of submissions, and to avoid bias towards the test set, it was decided to send only the results of the best model. The results for the HECKTOR Grand Challenge test set compared to the results of challenge participants are shown in Table~\ref{tab:hecktor_comparison}. Exemplary visualizations of the obtained results are shown in Figure~\ref{fig:hecktor_viz}.

\begin{table}[!h]
\centering
\caption{ Comparison of the results achieved by our V-Net with the results of the challenge participants who were included in the official ranking. The results of the challenge participants are presented in order of ranking and come from \cite{andrearczyk2022hecktor}. The results achieved by the V-Net are highlighted in bold.}
\begin{tabularx}{\linewidth}{C C C C}
\toprule
Team & \textbf{DSC\textsubscript{agg}GTVp ($\uparrow$)} & \textbf{DSC\textsubscript{agg}GTVn ($\uparrow$)} & \textbf{mean DSC\textsubscript{agg}($\uparrow$)} \\
\midrule
NVAUTO & $0.80066$ & $0.77539$ & $0.78802$ \\
SJTU426 & $0.77960$ & $0.77604$ & $0.77782$ \\
NeuralRad & $0.77485$ & $0.76938$ & $0.77212$ \\
LITO & $0.77700$ & $0.76269$ & $0.76984$ \\
TheDLab & $0.77447$ & $0.75865$ & $0.76656$ \\
MAIA & $0.75738$ & $0.77114$ & $0.76426$ \\
AIRT & $0.76689$ & $0.73392$ & $0.75040$ \\
AIMers & $0.73738$ & $0.73431$ & $0.73584$ \\
SMIAL & $0.68084$ & $0.75098$ & $0.71591$ \\
Test & $0.74499$ & $0.68618$ & $0.71559$ \\
BDAV USYD & $0.76136$ & $0.65927$ & $0.71031$ \\
\textbf{Our V-Net} & $\textbf{0.75491}$ & $\textbf{0.65396}$ & $\textbf{0.70443}$ \\
junma & $0.70906$ & $0.69948$ & $0.70427$ \\
RokieLab & $0.70131$ & $0.70100$ & $0.70115$ \\
LMU & $0.74460$ & $0.65610$ & $0.70035$ \\
TECVICO Corp & $0.74586$ & $0.65069$ & $0.69827$ \\
RT UMCG & $0.73741$ & $0.65059$ & $0.69400$ \\
HPCAS & $0.69786$ & $0.66730$ & $0.68258$ \\
ALaGreca & $0.72329$ & $0.61341$ & $0.66835$ \\
Qurit & $0.69553$ & $0.57343$ & $0.63448$ \\
VokCow & $0.59424$ & $0.54988$ & $0.57206$ \\
MLC & $0.46587$ & $0.53574$ & $0.50080$ \\
M\&H lab NU & $0.51342$ & $0.46557$ & $0.48949$ \\
\bottomrule
\end{tabularx}
\label{tab:hecktor_comparison}
\end{table}



\begin{figure}[!h]
  \centering
  \begin{minipage}{0.5\linewidth}
    \includegraphics[width=.9\linewidth]{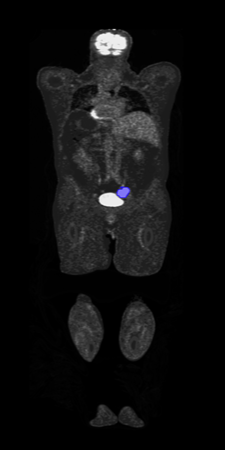}
    \label{fig:obrazek1}
  \end{minipage}%
  \hfill
  \begin{minipage}{0.5\linewidth}
    \includegraphics[width=.9\linewidth]{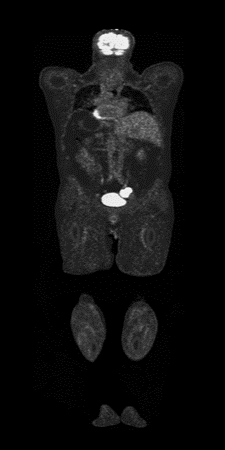}
    \label{fig:obrazek3}
  \end{minipage}
  \caption{An example of a false negative prediction case for the U-Net one-step segmentation where the model failed to detect the presence of a tumor (marked in blue in the ground truth) near the bladder.}
  \label{fig:fn_autopet}
\end{figure}

\section{Discussion}

Based on the results of the AutoPET dataset shown in Table 4.1, it can be concluded that removing tumor-free images during training improves the overall segmentation performance. Depending on the technique employed in training the model for positively annotated images, the resulting Dice coefficients varied between 0.39 and 0.69. Initially, it was considered that negative cases would notably reduce the quality of segmentation, indicating that outcomes for data containing cancer would be superior. The results obtained and statistical analysis confirmed this statement. For a dataset consisting solely of images with cancer lesions, the results ranged from 0.51 to 0.74. When working with a dataset of solely cancer lesions, the model can process more focused information that’s pertinent to tumor segmentation. This allows the model to more readily grasp crucial features for identifying cancer. 
In the case of two-stage segmentation, the initial segmentation was utilized to establish a prediction, which was then incorporated into the model as an additional input channel. Table 4.1 indicates a positive effect on the outcomes when compared to the first segmentation phase. The results in the second step are significantly better than those from the initial segmentation. The Dice coefficient value for U-Net was 0.58 initially but increased to almost 0.60 after additional segmentation. A corresponding effect was observed on the aggregated Dice, which yielded results of around 0.64 initially, and then increased to roughly 0.73. This is an expected effect because adding the initial segmentation results as an additional channel provides new contextual information to the network, which can help the network focus on improving relevant areas that were previously misclassified. The additional channel can act as a kind of regularization. The model endeavors to adapt itself to new information, thus fixating its focus on new data without any significant deviation from it. A certain limitation is introduced in the range of possible outputs that the next network can generate. Therefore, a fusion of the two methods may enhance segmentation accuracy. An added benefit was the dual-training of the network; the first segmentation phase was trained via classical methods, while the second phase utilized a patch-based approach, resulting in a minor increase of the required computational resources.

For the HECKTOR dataset, as shown in Table 2, the GTVn scores were notably lower (about 7-10\%) than the GTVp scores for all segmentation models. A possible explanation for this is that GTVp was present in a significantly larger number of images compared to GTVn, potentially making it more difficult for the model to learn the features of GTVn and leading to an increased number of false positives and false negatives.

The participants in the HECKTOR Grand Challenge 2022 achieved an aggregated Dice Similarity Coefficient for segmentation within the range of 0.46587-0.80066 for GTVp and 0.46557-0.77604 for GTVn, as shown in Table 2. Results obtained by the V-Net model were also within these ranges for the HECKTOR challenge test set. Although the presented results of other models in this thesis fall within these ranges, it is worth emphasizing that this comparison is only indicative. It should be noted that the models’ performance was evaluated only on a test set separate from the training set since no ground truths were available for the test set. Hence, it is uncertain whether the results of the models presented in this study would be equivalent if tested on the HECKTOR test set. The V-Net model achieved slightly worse results on the HECKTOR test set than on the test set that was separated from the training set, on which all models were evaluated and compared. It can be inferred that other models would behave similarly. Nearly half of the cases in the HECKTOR test set originated from centers that were excluded from the training set. This could have caused the model to have more trouble generalizing to data it had not previously encountered. Each center’s data may contain specific features that the model could not learn if it did not have access to the data. Furthermore, since the HECKTOR test set was significantly larger, there is an increased likelihood of errors, particularly false positives and false negatives. Nonetheless, despite these challenges, the model performed well since the variation in results was not very large.

For both datasets, the segmentation process encountered the greatest difficulties with the metabolically active areas surrounding or located near the analyzed structure. Such regions were inaccurately recognized by the algorithm as parts of segmented tumors. There have also been cases where a metabolically active structure near the tumor resulted in the tumor not being detected. There were cases where the tumor was not detected due to the presence of a metabolically active structure near the tumor. Such a scenario is illustrated in Figure~\ref{fig:fn_autopet} where a cancerous growth was situated close to a healthy, metabolically-active organ (bladder). Despite the lesion being sizeable and noticeable, the model failed to correctly identify it as a tumor, instead classifying it entirely as healthy tissue. In the case of the HECKTOR dataset, metabolically active structures in the neck were also problematic for the model and were misinterpreted by the algorithm as cancerous areas. Detecting cancer in tumor areas with low metabolic activity posed another challenge to the model. Metabolically active areas stand out especially on PET images due to their intense brightness. Another factor that may have had an impact on the results was the fact that the tumors in some of the images were very small. In such situations, a difference of just a few voxels between the prediction and the segmentation mask could significantly reduce the results. 

The study primarily analyzed how the choice of neural network architecture affected the obtained results. As such, the models were trained using similar parameters, and the training images were similarly prepared. It is worth noting that data preparation and training methods are also key factors in the effectiveness of deep learning models, in addition to selecting the appropriate model. The models utilized have the potential to yield better results if the data is properly prepared and optimal training parameters and techniques are chosen. This is confirmed by the results obtained by nnUNet using the AutoPET dataset. The results are significantly better with the changes introduces mostly to the initial preprocessing, augmentation, and postprocessing. It shows the crucial role of data preparation, pre- and post-processing, in contrast to the influence of the deep network architecture.

The size of the dataset also plays a crucial role. One of the main challenges in the automatic analysis of medical images is posed by small datasets. Deep learning models frequently perform better when trained on large datasets, thus expanding the dataset has the potential to enhance results. In the context of the size of the dataset, it is worth mentioning the UNETR model, which is based on the transformer architecture. Transformers have recently gained popularity and have become a new standard in many areas. In the context of medical image analysis, there are also many studies using transformers models and presenting promising results. A model utilizing the transformer architecture was assessed in this study, however, it did not yield superior outcomes when compared to convolutional neural networks. It is likely that the limited size of the dataset hindered the transformer’s ability to perform effectively. Transformers necessitate vast quantities of data to attain optimal efficiency. Potentially a self-supervsied pretraining approach could improve the UNETR results, however, this is within the scope of future work.

\section{Conclusions}

This study details several methods for tumor segmentation using PET/CT images. Only a few of the many available segmentation approaches have been utilized in the conducted experiments, suggesting that additional analysis may lead to better results. This is crucial in the context of implementing deep learning models in clinical practice, where precision and accuracy are indispensable for correct diagnoses and optimal treatment planning. The obtained results are comparable with the results reported in prior studies on the subject. While there is room for improvement, the results at this stage demonstrate the immense potential of deep learning models in the acceleration and automation of complex medical analysis.  Continued research in this area is necessary to further improve the effectiveness of diagnosis. Further work should focus on developing more advanced segmentation techniques and research on their integration with existing diagnostic systems. Nevertheless, the key takeaways from the results are connected with the importance of initial data preparation, augmentation and pre-processing that are more important that the deep architecture itself.

\section*{Acknowledgements}

We gratefully acknowledge Poland’s high-performance computing infrastructure PLGrid (HPC Centers: ACK Cyfronet AGH) for providing computer facilities and support within computational grant no. PLG/2023/016239.

\FloatBarrier

\bibliographystyle{abbrv}
\bibliography{refs}

\begin{thebibliography}{10}

\bibitem{nnUNetUNETR2023}
\url{https://github.com/MIC-DKFZ/nnUNetUNETR}.
\newblock Accessed: 21.08.2023.

\bibitem{hecktorwebpage}
\url{https://hecktor.grand-challenge.org/Data/ }Accessed: 2023-12-03.

\bibitem{evaluation}
\url{https://github.com/voreille/hecktor/blob/master/notebooks/evaluate_segmentation2022.ipynb}.
\newblock Accessed: 16.08.2023.

\bibitem{andrearczyk2022hecktor}
V.~Andrearczyk, V.~Oreiller, M.~Abobakr, A.~Akhavanallaf, P.~Balermpas, S.~Boughdad, L.~Capriotti, J.~Castelli, C.~Cheze Le~Rest, P.~Decazes, R.~Correia, D.~El-Habashy, H.~Elhalawani, C.~Fuller, M.~Jreige, Y.~Khamis, A.~Greca, A.~Mohamed, M.~Naser, and A.~Depeursinge.
\newblock Overview of the hecktor challenge at miccai 2022: Automatic head and neck tumor segmentation and outcome prediction in pet/ct.
\newblock volume 13626, pages 1--30. 03 2023.

\bibitem{andrearczyk2021overview}
V.~Andrearczyk, V.~Oreiller, S.~Boughdad, C.~C.~L. Rest, H.~Elhalawani, M.~Jreige, J.~O. Prior, M.~Valli{\`e}res, D.~Visvikis, M.~Hatt, et~al.
\newblock Overview of the hecktor challenge at miccai 2021: automatic head and neck tumor segmentation and outcome prediction in pet/ct images.
\newblock In {\em 3D head and neck tumor segmentation in PET/CT challenge}, pages 1--37. Springer, 2021.

\bibitem{andrearczyk2020overview}
V.~Andrearczyk, V.~Oreiller, M.~Jreige, M.~Vallieres, J.~Castelli, H.~Elhalawani, S.~Boughdad, J.~O. Prior, and A.~Depeursinge.
\newblock Overview of the hecktor challenge at miccai 2020: automatic head and neck tumor segmentation in pet/ct.
\newblock In {\em Head and Neck Tumor Segmentation: First Challenge, HECKTOR 2020, Held in Conjunction with MICCAI 2020, Lima, Peru, October 4, 2020, Proceedings 1}, pages 1--21. Springer, 2021.

\bibitem{azad2022medical}
R.~Azad, E.~K. Aghdam, A.~Rauland, Y.~Jia, A.~H. Avval, A.~Bozorgpour, S.~Karimijafarbigloo, J.~P. Cohen, E.~Adeli, and D.~Merhof.
\newblock Medical image segmentation review: The success of u-net.
\newblock {\em arXiv preprint arXiv:2211.14830}, 2022.

\bibitem{bauer2012automated}
C.~Bauer, S.~Sun, W.~Sun, J.~Otis, A.~Wallace, B.~J. Smith, J.~J. Sunderland, M.~M. Graham, M.~Sonka, J.~M. Buatti, et~al.
\newblock Automated measurement of uptake in cerebellum, liver, and aortic arch in full-body fdg pet/ct scans.
\newblock {\em Medical physics}, 39(6Part1):3112--3123, 2012.

\bibitem{bray2021ever}
F.~Bray, M.~Laversanne, E.~Weiderpass, and I.~Soerjomataram.
\newblock The ever-increasing importance of cancer as a leading cause of premature death worldwide.
\newblock {\em Cancer}, 127(16):3029--3030, 2021.

\bibitem{chow2020head}
L.~Q. Chow.
\newblock Head and neck cancer.
\newblock {\em New England Journal of Medicine}, 382(1):60--72, 2020.

\bibitem{monai_unetr}
M.~Contributors.
\newblock Unetr implementation in monai, 2023.
\newblock Accessed: 21.08.2023.

\bibitem{conze2023current}
P.-H. Conze, G.~Andrade-Miranda, V.~K. Singh, V.~Jaouen, and D.~Visvikis.
\newblock Current and emerging trends in medical image segmentation with deep learning.
\newblock {\em IEEE Transactions on Radiation and Plasma Medical Sciences}, 2023.

\bibitem{efron1994introduction}
B.~Efron and R.~J. Tibshirani.
\newblock {\em An introduction to the bootstrap}.
\newblock CRC press, 1994.

\bibitem{franzese2023enhancing}
C.~Franzese, D.~Dei, N.~Lambri, M.~A. Teriaca, M.~Badalamenti, L.~Crespi, S.~Tomatis, D.~Loiacono, P.~Mancosu, and M.~Scorsetti.
\newblock Enhancing radiotherapy workflow for head and neck cancer with artificial intelligence: A systematic review.
\newblock {\em Journal of Personalized Medicine}, 13(6):946, 2023.

\bibitem{gatidis2023autopet}
S.~Gatidis, M.~Fr{\"u}h, M.~Fabritius, S.~Gu, K.~Nikolaou, C.~La~Foug{\`e}re, J.~Ye, J.~He, Y.~Peng, L.~Bi, et~al.
\newblock The autopet challenge: Towards fully automated lesion segmentation in oncologic pet/ct imaging.
\newblock 2023.

\bibitem{gatidis2022whole}
S.~Gatidis, T.~Hepp, M.~Fr{\"u}h, C.~La~Foug{\`e}re, K.~Nikolaou, C.~Pfannenberg, B.~Sch{\"o}lkopf, T.~K{\"u}stner, C.~Cyran, and D.~Rubin.
\newblock A whole-body fdg-pet/ct dataset with manually annotated tumor lesions.
\newblock {\em Scientific Data}, 9(1):601, 2022.

\bibitem{illimoottil2023recent}
M.~Illimoottil and D.~Ginat.
\newblock Recent advances in deep learning and medical imaging for head and neck cancer treatment: Mri, ct, and pet scans.
\newblock {\em Cancers}, 15(13):3267, 2023.

\bibitem{isensee2021nnu}
F.~Isensee, P.~F. Jaeger, S.~A. Kohl, J.~Petersen, and K.~H. Maier-Hein.
\newblock nnu-net: a self-configuring method for deep learning-based biomedical image segmentation.
\newblock {\em Nature methods}, 18(2):203--211, 2021.

\bibitem{jiang2023review}
H.~Jiang, Z.~Diao, T.~Shi, Y.~Zhou, F.~Wang, W.~Hu, X.~Zhu, S.~Luo, G.~Tong, and Y.-D. Yao.
\newblock A review of deep learning-based multiple-lesion recognition from medical images: classification, detection and segmentation.
\newblock {\em Computers in Biology and Medicine}, page 106726, 2023.

\bibitem{li2022application}
S.~Li, J.~Liu, Z.~Wang, Z.~Cao, Y.~Yang, B.~Wang, S.~Xu, L.~Lu, M.~Iqbal~Saripan, X.~Zhang, et~al.
\newblock Application of pet/ct-based deep learning radiomics in head and neck cancer prognosis: a systematic review.
\newblock {\em Radiology Science}, 2022.

\bibitem{malhotra2022deep}
P.~Malhotra, S.~Gupta, D.~Koundal, A.~Zaguia, W.~Enbeyle, et~al.
\newblock Deep neural networks for medical image segmentation.
\newblock {\em Journal of Healthcare Engineering}, 2022, 2022.

\bibitem{mccrum2008correct}
E.~McCrum-Gardner.
\newblock Which is the correct statistical test to use?
\newblock {\em British Journal of Oral and Maxillofacial Surgery}, 46(1):38--41, 2008.

\bibitem{milletari2016v}
F.~Milletari, N.~Navab, and S.-A. Ahmadi.
\newblock V-net: Fully convolutional neural networks for volumetric medical image segmentation.
\newblock In {\em 2016 fourth international conference on 3D vision (3DV)}, pages 565--571. Ieee, 2016.

\bibitem{mo2022review}
Y.~Mo, Y.~Wu, X.~Yang, F.~Liu, and Y.~Liao.
\newblock Review the state-of-the-art technologies of semantic segmentation based on deep learning.
\newblock {\em Neurocomputing}, 493:626--646, 2022.

\bibitem{munir2019cancer}
K.~Munir, H.~Elahi, A.~Ayub, F.~Frezza, and A.~Rizzi.
\newblock Cancer diagnosis using deep learning: a bibliographic review.
\newblock {\em Cancers}, 11(9):1235, 2019.

\bibitem{noguchi2011bootstrap}
K.~Noguchi, Y.~R. Gel, and C.~R. Duguay.
\newblock Bootstrap-based tests for trends in hydrological time series, with application to ice phenology data.
\newblock {\em Journal of Hydrology}, 410(3-4):150--161, 2011.

\bibitem{oreiller2022head}
V.~Oreiller, V.~Andrearczyk, M.~Jreige, S.~Boughdad, H.~Elhalawani, J.~Castelli, M.~Vallieres, S.~Zhu, J.~Xie, Y.~Peng, et~al.
\newblock Head and neck tumor segmentation in pet/ct: the hecktor challenge.
\newblock {\em Medical image analysis}, 77:102336, 2022.

\bibitem{perez2021torchio}
F.~P{\'e}rez-Garc{\'\i}a, R.~Sparks, and S.~Ourselin.
\newblock Torchio: a python library for efficient loading, preprocessing, augmentation and patch-based sampling of medical images in deep learning.
\newblock {\em Computer Methods and Programs in Biomedicine}, 208:106236, 2021.

\bibitem{rebaud2022simplicity}
L.~Rebaud, T.~Escobar, F.~Khalid, K.~Girum, and I.~Buvat.
\newblock Simplicity is all you need: out-of-the-box nnunet followed by binary-weighted radiomic model for segmentation and outcome prediction in head and neck pet/ct.
\newblock In {\em 3D Head and Neck Tumor Segmentation in PET/CT Challenge}, pages 121--134. Springer, 2022.

\bibitem{ronneberger2015u}
O.~Ronneberger, P.~Fischer, and T.~Brox.
\newblock U-net: Convolutional networks for biomedical image segmentation.
\newblock In {\em Medical Image Computing and Computer-Assisted Intervention--MICCAI 2015: 18th International Conference, Munich, Germany, October 5-9, 2015, Proceedings, Part III 18}, pages 234--241. Springer, 2015.

\bibitem{shi2022deep}
F.~Shi, W.~Hu, J.~Wu, M.~Han, J.~Wang, W.~Zhang, Q.~Zhou, J.~Zhou, Y.~Wei, Y.~Shao, et~al.
\newblock Deep learning empowered volume delineation of whole-body organs-at-risk for accelerated radiotherapy.
\newblock {\em Nature Communications}, 13(1):6566, 2022.

\bibitem{shiri2024information}
I.~Shiri, M.~Amini, F.~Yousefirizi, A.~Vafaei~Sadr, G.~Hajianfar, Y.~Salimi, Z.~Mansouri, E.~Jenabi, M.~Maghsudi, I.~Mainta, et~al.
\newblock Information fusion for fully automated segmentation of head and neck tumors from pet and ct images.
\newblock {\em Medical Physics}, 51(1):319--333, 2024.

\bibitem{thisanke2023semantic}
H.~Thisanke, C.~Deshan, K.~Chamith, S.~Seneviratne, R.~Vidanaarachchi, and D.~Herath.
\newblock Semantic segmentation using vision transformers: A survey.
\newblock {\em Engineering Applications of Artificial Intelligence}, 126:106669, 2023.

\bibitem{tran2021deep}
K.~A. Tran, O.~Kondrashova, A.~Bradley, E.~D. Williams, J.~V. Pearson, and N.~Waddell.
\newblock Deep learning in cancer diagnosis, prognosis and treatment selection.
\newblock {\em Genome Medicine}, 13(1):1--17, 2021.

\bibitem{wang2021deep}
X.~Wang and B.-b. Li.
\newblock Deep learning in head and neck tumor multiomics diagnosis and analysis: review of the literature.
\newblock {\em Frontiers in Genetics}, 12:624820, 2021.

\bibitem{wang2024comparison}
Y.~Wang, E.~Lombardo, L.~Huang, M.~Avanzo, G.~Fanetti, G.~Franchin, S.~Zschaeck, J.~Weing{\"a}rtner, C.~Belka, M.~Riboldi, et~al.
\newblock Comparison of deep learning networks for fully automated head and neck tumor delineation on multi-centric pet/ct images.
\newblock {\em Radiation Oncology}, 19(1):3, 2024.

\bibitem{xiao2023transformers}
H.~Xiao, L.~Li, Q.~Liu, X.~Zhu, and Q.~Zhang.
\newblock Transformers in medical image segmentation: A review.
\newblock {\em Biomedical Signal Processing and Control}, 84:104791, 2023.

\bibitem{xue2024multi}
H.~Xue, Y.~Yao, and Y.~Teng.
\newblock Multi-modal tumor segmentation methods based on deep learning: a narrative review.
\newblock {\em Quantitative Imaging in Medicine and Surgery}, 14(1):1122, 2024.

\bibitem{zhou2017deep}
X.~Zhou, R.~Takayama, S.~Wang, T.~Hara, and H.~Fujita.
\newblock Deep learning of the sectional appearances of 3d ct images for anatomical structure segmentation based on an fcn voting method.
\newblock {\em Medical physics}, 44(10):5221--5233, 2017.

\end{thebibliography}
\end{document}